\documentstyle[aps]{revtex}
\date{23 feb 2001}

\begin{document}

\title {Coherent Radiation of the Crystalline Beam}

\author {L.A.Gevorgian,R.V. Tumanian\\
Yerevan Physics Institute}

\maketitle
\begin{abstract}

The coherent radiation of the ordered or crystalline electron beam
is investigated. For the first time it is shown that crystallization
conditions for charged particles beams differs from those of plasma.
Crystallization conditions for charged relativistic bunches are found
and shown that this conditions satisfies in some existing linacs.The
coherent undulator radiation of crystalline beams in the XUV region
are considered.Necessary conditions of coherency are found. The influence of
deviations from mean distance on the radiation coherency is considered.
\end{abstract}
\begin{twocolumn}
\flushleft
\section{Introduction}

 The importance of tunable and powerful sources of the coherent
 radiation on XUV wavelenghts is cause in recent years few projects
 of FEL  \cite{1,2}. This FEL's operating in the SASE mode, i.e.
starting from noise in the initial electron beam longitudinal density
distribution. This evolution determines the undulatory length needed
to reach saturation practically very long (about hundred meters). The
needed power of laser radiation is possible to obtain in other way. In
last two decades very important  experimental \cite{9,10}and theoretical
\cite{11-14} results are achieved in investigations of the ordered
bunches or crystalline beams.The possibility of such new and interesting
state of matter is become real in the storage rings of charged particles
and as we show below in linear accelerators with high density beams.
The case of real storage ring lattice was considered in \cite {15}.
The crystalline or ordered beams have of course many interesting and
important properties and applications.Few properties and one very
important application is considered in this paper. It is clear that
particles of dense bunches are arranged in the certain orders.In these
ordered bunches particles are complete transverse planes,which are same
spacing in the longitudinal direction.The conditions and properties of
such ordering are considering in the next section of this paper.Such
bunches are radiate coherent at the wavelengths integer times smaller
than inter plane distances, as shown in the section 3.This radiation
calls super radiant regime of FEL radiation \cite{3-7} or coherent
spontaneous emission {CSE}.As shown in
above references it is possible in two cases.First,for short bunches
and second for modulated bunches. The super radiant regime because of
self bunching of the beam in the FEL is considered by Bonifacio \cite{6,7}.
In difference from above references where is considered the coherency
of long wavelength radiation (the wavelength much more than mean
distance between bunch electrons), in this report is considered the
coherent radiation of the bunch when the radiation wavelength is about
or less than distance between particles. In this case it is important
the discreteness of beam and correlations between beam particles
positions,because of strong Coulomb interaction of the beam particles.

\section{Ordered or Crystalline beams}
 The requirement of bunch uniformity  assumes that electrons
of bunch with density n are replaced on the same mean distance
$\bar{a}=n^{-1/3}$ from each other.Such replacement is possible
 only when each three particles are compose equilateral triangle
as result of strong correlation between particles positions.This
is connected with approximately hexagonal structure of the disordered
medium \cite{8} at most probable.Each particle in such medium have
14 nearest neighbors,but numerical calculations show that the mean
number of nearest neighbors is about 15 with mean deviation 10 percents.
For medium with long distance inter particle interaction such as
beam or bunch the triangle apices are the equilibrium points of the
particles positions.The particles which are not replaced in his
equilibrium points are oscillate around his equilibrium point with
frequency $\Omega=N_0\omega _p$,where $\omega _p=\sqrt{4\pi e^{2}n/mc}$
plasma frequency ,$N_0$-number of particles in the transverse
direction, and amplitude equal to deviation from equilibrium
point.It is well known that properties of any medium are depend on
dimensionless parameter $\Gamma$,which is the ratio of the depth of
the interparticle potential well and medium temperature. For neutral
and one component plasma (OCP) because of Debay screening inter particle
potential is about pure two particle Coulomb potential $e^2/\bar a$
\cite{13}.The numerical Molecular Dynamics (MD) calculations show that
for $\Gamma\geq 172$ OCP is crystallize with body centered cube (bcc)
lattice \cite{12}.Detailed MD calculations \cite{11} for finite full
charged Coulomb systems with number of particles about thousand was
show that in this case crystallization take place at low values of
$\Gamma$,but this result is not explained theoretically.We show in this
report that for full charged Coulomb systems the potential well of each
particle is about $N_0$ times deeper than pure two particle Coulomb potential.
 After expanding of the full force between particles with charge e and
fixed distance R moving along longitudinal z direction with velocity
$v=\beta c$
\begin{eqnarray}
F_z=\frac{e^2}{R^2}\frac{(1-\beta^2)cos{\theta}}{(1-\beta^{2}(sin{\theta})^2)^3/2}\\
F_\perp=\frac{e^2}{R^2}\frac{(1-\beta^2)sin{\theta}}{(1-\beta^{2}sin^2{\theta})^{3/2}}
\end{eqnarray}
where $\theta$-angle between R and z directions, around equilibrium
points of the bcc lattice,one can find that potential wells in the
longitudinal and transverse directions are
\begin{eqnarray}
U_\|=U_{0}\frac{N_0}{4\gamma^2}S_l&,&S_l=\sum{\frac{1}{n^3}}\\
U_\bot=U_0\frac{\lambda _l}{\lambda _t}\frac{\gamma}{4} S_t &,&S_t=\sum{\frac{n_1}{(n_1^2+n_2^2)^2}}
\end{eqnarray}

where $U_0=e^2/\lambda _l$ bare Coulomb potential.The sum in the $S_l$ is
carrying out over number of transverse planes,and the sum in the $S_t$ over
particles of one plane.Here is assumed that self mean space charge forces
of the bunch are compensated by external restoring forces \cite{11,14}.
Such deepening of the potential well in the bunches in comparison with
known potential well in the OCP is mean that crystallization or ordering
of bunch is possible at comparable higher temperatures than of OCP.Notice,
that transverse ordering is more difficult than longitudinal ordering,
because of small tranverse well.So,we can neglect the transverse potential
and consider the bunch as consisting of transverse planes with spacing,
but particles on this planes are replaced randomly. Particles of the
bunch in this longitudinal potential well is oscillate with frequency

$$\Omega ^2=\frac{U}{2m\lambda ^2} $$

where $\lambda$ -mean interparticle distance in correspondence direction,
m is effective mass of bunch particle equal to $m_0\gamma ^3$ for
linacs and $m_0\gamma ^3/(1-\alpha\gamma ^2)$ for cyclic accelerators.
If the crystallization time which equal few oscillation periods $l_c=c/\Omega$
much less than period of betatron oscillations of the accelerator the
bunch becomes ordered.The complete consideration show that adiabatic
damping of the bunch sizes after compression during acceleration
equavalent to decreasing of the $\Omega$,which is means that
oscillation energy of the particle in the well is decreasing.
The electron bunch with $\lambda$ about 0.1 nm,
and $\gamma=10^4$ (SLAC)have $l_c$ is about 16 cm and shorter than betatron
oscillation wavelength.The calculations show that beam of the SLAC in
the Final Focus have $\Gamma$ about few hundreds which is means that
bunch may be crystallized.

\section{Coherent Radiation of the Bunch}

Radiation of the ensemble of N  electrons moving along identical
trajectories but with arbitrary spatial displacement may be written
in the form \cite{3}
\begin{equation}
I=iNF
\end{equation}
where i-intensity of single electron radiation,N-number of the
electrons in the bunch,F-factor of coherency of the bunch and may
be written in the follow form \cite{3}

$$F=\frac{1}{N}\sum{e^{i\vec{k}\vec{r_j}}}\sum{e^{-i\vec{k}\vec{r_j}}}$$
The position of the j-th electron $\vec{r_j}$ may be written as a
sum of transverse and longitudinal parts $\vec{r_j}=\vec{r_jtr}+z_j$.
For waves in the longitudinal direction the transverse part of the F is equal
to $N_r^2$.This is right because of transverse coherency of the radiation \cite{3}.
For $N_z$ planes with $N_r$ electrons on each plane the coherency factor is
may be written in the following form
$$F=\frac{N_{r}^2}{N}\sum{e^{ikj\tilde{a}}}\sum{e^{-ikj\tilde{a}}}$$

where  $\sum{}$ -means sum over planes in longitudinal direction.
It is not difficult to obtain that
$$F=N_{r}^2\frac{\sin^2{N_{z}x}}{\sin^2{x}} $$
and when the value $x=k\tilde{a}$ is approach to $n\pi$ (resonance
condition) the coherency factor is becomes $N^2$.
This consideration is true for bunch with step function for
longitudinal density distribution (for homogeneous beam).
Now consider the influence of fluctuations or deviations of the
particles positions from hexagonal model above on the coherence short
wave length radiation of the bunch.Let us assume that particle is
on the mean distance, but not in the z-direction. In this case the
particle position is fluctuated on the value $\tilde{a}\theta^2/2$, and
if $x\theta^2/4\ll 1$ we may neglect this fluctuation.Notice,
that in important practical cases this condition is satisfied. It
is not satisfied only for very short radiation when $x\gg 1$.
The second source of fluctuations from our model is possible when
the particle is replaced on the z-direction but not in the
distance a. In this case F multiplied by the factor
$e^{-k^{2}b^2}$, where b- the dispersion of the fluctuation. It is
clear that if $b/\lambda\ll 1$, this fluctuations not disturb the
radiation coherency.Executed numerical calculations show that
fluctuations about ten times smaller than inter particle distance a.
This show that radiation of the wavelengths few times shorter than
interparticle distance is availably too.Notice,that such full coherent
radiation is much powerful than any other radiation regime in the same
conditions.

\section{Conclusions}
So,in this paper are found the crystallization or ordering conditions
of dense bunches or beams,which may be satisfied much easier than those
for OCP.This very exotic and interesting station of the matter can be
obtained at existing linac beams.This possibility is very important
for attaining crystalline state of the beams as well as for many other 
applications of new state of matter.We show that ordered bunches can 
radiate coherently ,i.e. much powerful than spontaneous one or SASE.
In dependence of beam energy and density this radiation may have 
wavelength at XUV region also.

\end{twocolumn}
\end{document}